\documentclass[twocolumn,prb,aps,amsmath,amssymb,superscriptaddress,showpacs]{revtex4}
\usepackage{graphicx}
\usepackage{bm}
\usepackage[usenames,dvipsnames]{color}

\newcommand{\telque}{;}
\newcommand\id{\mathrm{Id}}
\newcommand\calA{\mathcal{A}}

\newcommand\calE{\mathcal{E}}
\newcommand\calH{\mathcal{H}}
\newcommand\calJ{\mathcal{J}}
\newcommand\calK{\mathcal{K}}
\newcommand\calM{\mathcal{M}}

\newcommand\bfc{\mathbf{c}}
\newcommand\bfd{\mathbf{d}}
\newcommand{\un}{\mathbb{I}}
\newcommand{\pink}[1]{#1}
\newcommand{\bleu}[1]{#1}

\begin{document}
\title{The Standard Model as an extension of the
noncommutative algebra of forms},

\author{Christian Brouder}
 \affiliation{Institut de Min\'eralogie, de Physique des Mat\'eriaux et de
 Cosmochimie, \\ Sorbonne Universit\'es, UMR CNRS 7590, 
 Universit\'e Pierre et Marie Curie-Paris 06,
 \\ Mus\'eum National d'Histoire Naturelle, IRD UMR 206,
 4 place Jussieu, F-75005 Paris, France}
\author{Nadir Bizi}
 \affiliation{Institut de Min\'eralogie, de Physique des Mat\'eriaux et de
 Cosmochimie, \\ Sorbonne Universit\'es, UMR CNRS 7590, 
 Universit\'e Pierre et Marie Curie-Paris 06,
 \\ Mus\'eum National d'Histoire Naturelle, IRD UMR 206,
 4 place Jussieu, F-75005 Paris, France}
\author{Fabien Besnard}
 \affiliation{P{\^o}le de recherche M.L. Paris,
 EPF, 3~bis rue Lakanal,
 F-92330 Sceaux, France}
\date{\today}
\begin{abstract}
The Standard Model of particle physics can be deduced
from a small number of axioms within Connes' noncommutative
geometry (NCG). Boyle and Farnsworth [\emph{New J. Phys.}
{\bf{16}} (2014) 123027] proposed to interpret
Connes' approach as an algebra extension in the sense
of Eilenberg. By doing so, they could deduce three
axioms of the NCG Standard Model 
(i.e. order zero, order one and massless photon)
from the single requirement that the extended algebra be
associative. However, their approach was only applied to the finite
algebra \pink{and fails the full model}.

By taking into account the differential graded structure
of the algebra of noncommutative differential forms, 
we obtain a formulation where the same three axioms
are deduced from the associativity of the extended
differential graded algebra, but which is now \pink{also} compatible with 
the full Standard Model.

\pink{Finally, we present a Lorentzian version of
the noncommutative geometry of the Standard Model and
we show that the three axioms} still hold
if the four-dimensional manifold has a Lorentzian metric.
\end{abstract}
\maketitle

\section{Introduction}

Noncommutative geometry provides a particularly elegant
way to derive and describe the structure and the Lagrangian
of the Standard Model in curved spacetime and
its coupling to gravitation~\cite{Connes-Marcolli}.
The main ingredients of this approach \pink{are}  an
algebra $\calA=C^\infty(M)\otimes \calA_F$
(where $M$ is a Riemann spin manifold and
$\calA_F=\mathbb{C}\oplus\mathbb{H}\oplus 
M_3(\mathbb{C})$),\cite{Connes-Marcolli}
a Hilbert space $\calH=L^2(M,S)\otimes \calH_F$
(where \pink{$S$ is the spinor bundle} and
$\calH_F$ is 96-dimensional), and a 
Dirac operator $D$.

The elements $a$  of the algebra are represented by
bounded operators $\pi(a)$ over $\calH$.
In this approach, 
the gauge bosons are described
by gauge potentials (i.e. noncommutative one-forms) in 
$\Omega^1_D \calA$,
where $\Omega_D \calA$ is the differential graded
algebra (DGA) constructed from $\calA$, whose differential
is calculated by using the commutator with $D$.

From the physical point of view, a striking
success of the noncommutative geometric approach is that
the algebra, the Hilbert space and the Dirac operator
of the Standard Model can be derived from a few
simple axioms, including the condition of order zero,
the condition of order one and the condition of massless 
photon.~\cite{Chamseddine-07,Chamseddine-08,Chamseddine-10}
Then, the Lagrangian of the Standard Model coupled
to (Riemannian) gravity is obtained by counting
the eigenvalues of the Dirac operator $D$.~\cite{Connes-Marcolli}

Still, this approach is not completely physical
because it is formulated in the Riemannian
(instead of Lorentzian) signature and is not
quantized. 
Therefore, the original NCG approach was 
variously modified, by using 
Lie algebras~\cite{Wulkenhaar-99},
twisted spectral triples~\cite{Connes-08,Devastato-15}
or Lie algebroids and derivation-based NCG,\cite{Lazzarini-12}
to deal with models that do not enter into the
standard NCG framework (e.g.  quantum groups
or Grand Symmetry).\cite{Devastato-14}

However, to discover the mathematical framework
\pink{most suitable for}
the physical Standard Model, 
it might \pink{also be useful to go the other way
and find an approach} where
the Standard Model is still more constrained
(i.e. deduced from less axioms) than in NCG.
Boyle and Farnsworth\cite{Boyle-14} recently
used Eilenberg's algebra extension method to build 
an algebra $\calE$ where the universal DGA
$\Omega$ built on $\calA$
(see section~\ref{ncdiform})
is extended by the  Hilbert space $\calH$.
This is physically more
satisfactory because the gauge field, the field intensity,
the curvature
and the Lagrangian densities are noncommutative differential forms,
which belong to $\Omega$ up to an ideal described below.
They observed that the associativity of the algebra
$\calE$ imposes a new condition (of order two) which
is satisfied by the finite part $\calA_F$ of the
Standard Model and removes
a somewhat arbitrary axiom in 
Chamseddine and Connes' derivation.\cite{Chamseddine-08}
This axiom requires the Dirac operator $D_F$
of the finite algebra to commute with a specific family
of elements of $\calA_F$.
It is called the condition of massless photon
because it ensures that the photon has no mass.

However, as noticed by Boyle and Farnsworth, this approach
has two drawbacks: i) it is not valid for a spin manifold
(i.e. the canonical spectral triple $(C^\infty(M),L^2(M,S),D_M)$
does not satisfy the condition of order two);
(ii) it uses the DGA algebra $\Omega$ in which
gauge fields with vanishing representation
(i.e. $A\in \Omega$ such that $\pi(A)=0$) 
can have non-zero field intensity
(i.e. $\pi(dA)\not=0$). This makes the 
Yang-Mills action ill defined.\cite{Connes-Marcolli}
A consistent substitute for $\Omega$ is
the space $\Omega_D$ of noncommutative differential forms
which is a DGA built as the quotient of $\Omega$ by
a differential ideal $J$ usually called the \emph{junk}.

To solve both problems, we define an extension $\calE$
of the physically meaningful  algebra $\Omega_D$ of 
noncommutative differential forms
by a representation space $\calM_D$ that we build explicitly.
Since the algebra $\Omega_D$ is a DGA,
we require the extension $\calE$ to be also a DGA
and we obtain that
$\calM_D$ must be a differential graded bimodule over $\Omega_D$
(see below).
The most conspicuous consequence of this construction is
a modification of the condition of order two proposed
by Boyle and Farnsworth, 
which provides
exactly the same constraints on the finite part of the
spectral triple of the Standard Model, but which is now
consistent with
the spectral triple of a spin manifold.
As a consequence,
the full spectral triple of the Standard Model
(and not only its finite part) now satisfies the condition
of order two and enables us to remove the condition
of massless photon.

In the next section, we describe the extension of an
algebra by a vector space, first proposed by 
Eilenberg~\cite{Eilenberg-48} and used by Boyle
and Farnsworth. We discuss the modification required
to take the differential graded structure into account.
Then, we describe the construction by Connes and Lott 
of noncommutative differential forms, we build a
space $\calM_D$ that can be used to extend 
$\Omega_D$ into a DGA $\calE$ \pink{and}
we show that the spectral triple of the
Standard Model fits into this framework if and only if the
condition of massless photon is satisfied.
\pink{Finally, we add a fundamentaly symmetry
that makes the model compatible with the 
Lorentz signature of the spin manifold $M$, thus
reaching the fully physical Standard Model.}

\section{Extension of algebras}

The cohomology of Lie algebras plays a crucial role in the modern 
understanding of classical~\cite{Giachetta} and quantum~\cite{Henneaux}
gauge field theories. This cohomology theory mixes the
gauge Lie algebra and its representation over a vector 
or spinor bundle.
Shortly after the publication of the cohomology of Lie algebras,
Eilenberg generalized this idea to the 
representation of any algebra whose product
is defined by a bilinear map with possible linear
constraints (Lie, associative, Jordan, commutative,
etc.).~\cite{Eilenberg-48}

\subsection{Eilenberg's extension}
\label{Eilsect}
If $A$ is a (possibly non-associative) algebra with product 
$ a\cdot b$ and $V$ is a vector space, then the
(possibly non-associative) algebra $(E,\star)$ is an extension
of $A$ by $V$ if $V\subset E$ and there is a linear map 
$\varphi:E\to A$ such that
$\varphi(e)=0$ iff $e\in V$,
$\varphi(e\star e')=\varphi(e)\cdot\varphi(e')$
for every $e$ and $e'$ in $E$ and 
$u \star v=0$ when $u$ and $v$ are in $V$.

Eilenberg showed that, if $\eta:A\to E$ is a 
map such that $\eta(a)$ represents $a$ in $E$
(i.e. for every $a\in A$, $\varphi(\eta(a))=a$), then 
$\eta(a\cdot b)=\eta(a)\star \eta(b)+f(a,b)$,
where $f$ is a bilinear map $A\times A \to V$.
The product $\star$ in $E$ induces two bilinear maps
$(a,v)\mapsto a\triangleright v=\eta(a)\star v$
and
$(v,a)\mapsto v\triangleleft a=v\star \eta(a)$
and it can be shown that $a\triangleright v$ and
$v\triangleleft a$ are in $V$ and independent of $\eta$.
Conversely, a product in $A$ and two blinear maps
$\triangleright$ and $\triangleleft$ determine 
an extension $E$ of $A$ by $V$ and a product $\star$, 
which are unique up to an equivalence determined by $f$.

Noncommutative geometry belongs to this framework if
we define $A=\calA$, $V=\calH$,
$a\triangleright v=\pi(a) v$, where $\pi(a)$ is
the representation of $a$ in the space $B(\calH)$
of bounded operators on $\calH$ and 
$v\triangleleft  b=\pi(b)^\circ  v$,
where $\pi(b)^\circ=J\pi(b)^\dagger J^{-1}$ and $J$ is an
antilinear isometry called the \emph{real structure}.
Notice that, in the last expression, the right
action $v\triangleleft  b$ of $b$ on $v$ is replaced
by the left product by  $\pi(b)^\circ$ on $v$.
This remark will turn out to be crucial.
When there is no ambiguity, we sometimes use a common
abuse of notation and write $a$ for $\pi(a)$ and
$b^\circ$ for $\pi(b)^\circ$.

Then, Eilenberg showed that the extension $E$ is associative 
iff the following conditions are satisfied:
\begin{eqnarray}
   a\cdot(b\cdot c) &=& (a\cdot b)\cdot c,\label{Eil1}\\
   a\triangleright (b\triangleright v) &=& 
   (a\cdot b)\triangleright v,\label{Eil2}\\
   (v\triangleleft a)\triangleleft b &=& 
   v\triangleleft(a\cdot b),\label{Eil3}\\
   (a\triangleright v)\triangleleft b &=& a\triangleright 
         (v\triangleleft b),\label{Eil4}\\
   a\triangleright f(b,c) + f(a,b\cdot c) &=& 
   f(a\cdot b,c) + f(a,b)\triangleleft c\label{Eil5}.
\end{eqnarray}
Condition (\ref{Eil1}) means that $A$ is an associative
algebra, condition (\ref{Eil2}) that
$\triangleright$ is a left action of $A$ on $V$,
condition (\ref{Eil3}) that
$\triangleleft$ is a right action of $A$ on $V$,
condition (\ref{Eil4}) that
the right and left actions are compatible
(i.e. that $V$ is a bimodule),
the map $f$ in
condition (\ref{Eil5}) is required for the extension
to have better functorial properties but we 
do not use it here and we consider the case
$E=A\oplus V$, $\varphi(a+v)=a$, $\eta=\id$ and $f=0$.
In the NCG example of the extension of $\calA$
by $\calH$ that we gave \pink{in the previous}
paragraph, condition (\ref{Eil4})
becomes $\pi(b)^\circ \pi(a)=\pi(a)\pi(b)^\circ$,
which is the condition of order zero of NCG usually written
$[a,b^\circ]=0$.

\subsection{Differential graded-representation of a DGA}
We noticed in the introduction that 
$\Omega$ and $\Omega_D$ are DGA.
It is now time to explain what that means.
A graded vector space is a direct sum
$V=\bigoplus_{n\ge0} V^n$ of vector spaces.
\pink{If $v\in V$ belongs to some $V^n$ we 
say that $v$ is \emph{homogeneous} and that its
\emph{degree} is}
$|v|=n$.
A DGA is a graded vector space $A$
equipped with an associative product $\cdot$ and a differential $\delta$. 
The product of \pink{the} algebra satisfies
$|a\cdot b|=|a|+|b|$. The differential satisfies
\pink{$|\delta a|=|a|+1$},
$\delta^2=0$ and the graded Leibniz rule
$\delta(a\cdot b)=(\delta a)\cdot b + (-1)^{|a|}a \cdot(\delta b)$.
Differential graded algebras are a basic tool of 
cohomological physics.\cite{Stasheff-88}

A graded left-representation of $A$ 
is a graded vector space $\calM$ with
a left action $\triangleright$ of $A$ over $\calM$
such that $|a\triangleright m|=|a|+|m|$, with a similar
definition for a graded right-representation.
A differential graded left-representation of $A$
is a graded left-representation $\calM$
furnished with a differential $\delta:\calM^n\to \calM^{n+1}$
such that:\cite{Neisendorfer}
\begin{eqnarray}
\delta(a\triangleright m) &=& (\delta a)\triangleright m
+(-1)^{|a]} a\triangleright (\delta m).
\label{deltaright}
\end{eqnarray}
Similarly, in a differential graded right-representation:
\begin{eqnarray}
\delta(m\triangleleft a) &=& (\delta m)\triangleleft a
+(-1)^{|m]} m\triangleleft (\delta a).
\label{deltaleft}
\end{eqnarray}
The signs in Eqs.~(\ref{deltaright}) and (\ref{deltaleft})
are imposed by the fact that an algebra is a
left and right representation of itself.
By a straightforward generalization of Eilenberg's result, we
see that $E$ is an extension of $A$ by $\calM$
as a DGA iff $\calM$
is a differential graded bimodule over $A$
(i.e. $\calM$ is a differential
graded left-representation, a differential graded
right-representation and the left and right actions
are compatible in the sense of Eq.(\ref{Eil4})).

We saw in section~\ref{Eilsect} that in the NCG
framework, the right action of an element $a$
of the algebra is represented as a left product
by the operator \pink{$\pi(a)^\circ$}. \pink{To} retain
this type of representation for $\calM$,
we define a
linear map $a\mapsto a^\circ$ such that the
right action by $a$ (i.e. $m\triangleleft a$) is represented by the
left product with $a^\circ$ (i.e. $a^\circ m$).
We do not assume that $a^\circ$ belongs to the algebra $A$
but we require that $|a^\circ|=|a|$.
However, the representation of $m\triangleleft a$
by $a^\circ m$ is \pink{different from the case where}
$A$ is only an algebra because
compatibility
with the DGA structure imposes 
the following sign:\cite{Neisendorfer}
\begin{eqnarray}
a^\circ \, m &=& (-1)^{|a||m|} m \triangleleft a.
\label{rightleft}
\end{eqnarray}
Indeed, $\delta(a^\circ m)=\delta(a^\circ) m
+(-1)^{|a|}a^\circ \delta m$ implies \pink{now}
\begin{eqnarray*}
(-1)^{|a||m|} \delta(m \triangleleft a) &=&
(-1)^{(|a|+1)|m|} m \triangleleft (\delta a)
\\&&
+ (-1)^{|a|+|a|(|m|+1)} (\delta m) \triangleleft a,
\end{eqnarray*}
and we recover Eq.~(\ref{deltaleft}).

The map ${}^\circ$ is compatible with the 
differential graded bimodule
structure if the following conditions hold
for every $a$ and $b$ in $A$:
\begin{eqnarray}
a^\circ \,b^\circ &=& (-1)^{|a||b|} (b\cdot a)^\circ,
\label{aoactbo}\\
a^\circ \,b &=& (-1)^{|a||b|} 
b \,a^\circ,
\label{aoactb}\\
\delta (a^\circ) &=& (\delta a)^\circ.
\label{deltaao}
\end{eqnarray}
Equation~(\ref{aoactbo}) follows from
Eqs.~(\ref{Eil3}) and (\ref{rightleft}),
Eq.~(\ref{aoactb}) follows from
Eqs.~(\ref{Eil4}) and (\ref{rightleft}).
To derive Eq.~(\ref{deltaao}), we apply 
transformation~(\ref{rightleft}) to Eq.~(\ref{deltaleft}) to
obtain
$\delta(a^\circ m)=(-1)^{|a|} a^\circ\delta m +(\delta a)^\circ m$
and we compare with the expression for
$\delta(a^\circ m)$ given after Eq.~(\ref{rightleft}).

\section{Noncommutative differential forms}
\label{ncdiform}
If $\calA$ is an algebra, its associated universal differential
graded algebra $\Omega=\bigoplus_{n\ge0} \Omega^n$ 
is defined as follows.~\cite{Connes94,Landi-NCG}
In degree zero $\Omega^0=\calA$. The space
$\Omega^n$ is generated by 
the elements $a_0 (\delta a_1)\dots(\delta a_n)$,
where $a_0, \dots, a_n$ are elements of $\calA$ and
$\delta$ is a linear operator satisfying $\delta^2=0$,
$\delta(a)=(\delta a)$, and
$\delta (\omega\rho)=(\delta \omega) \rho + (-1)^{|\omega|}
\omega(\delta\rho)$.

In NCG, $\delta a$ is represented over $\calH$ as the (bounded)
operator $[D,a]$ and the $n$-form
$\omega=a_0 (\delta a_1)\dots(\delta a_n)$ is represented
by $\pi(\omega)=\pi(a_0)[D,\pi(a_1)]\dots [D,\pi(a_n)]$
where $\pi$ becomes now a $*$-representation of $\Omega$. 
However, this representation is not graded
(a fact which is sometimes overlooked) because
$\pi(\Omega)\subset B(\calH)$\pink{, which is
not graded}. To obtain a graded
representation we
replace $\pi$ by $\tilde \pi : \Omega\rightarrow 
B^\infty(\calH)=\bigoplus_n V^n$ where each $V^n$
is $B(\calH)$: if $\omega\in \Omega^n$, then
$\tilde\pi(\omega)\in V^n$.
However, the difference between $\pi$ and $\tilde \pi$ is invisible as 
long as we only consider homogeneous elements and are careful 
about their degree. This is why we will stick to the notation $\pi$ 
in the sequel when no confusion can arise.

The representation $\pi$ (i.e. $\tilde{\pi}$) is now a graded
$*$-representation of $\Omega$ \pink{considered as
a graded algebra}.\cite{Connes94}
However,
it is not a well-defined representation of the
differential because
there can be $n$-forms $\omega$ such
that $\pi(\omega)=0$ and $\pi(\delta\omega)\not=0$,
as we illustrate now with
the spectral triple of a spin manifold.

Let $f$ and $g$ be two functions in $C^\infty(M)$. They
are represented \pink{by multiplication} over $\calH=C^\infty(M,S)$:
$\pi(f)\psi=f\psi$. Then, $\delta f$ is represented by
$\pi(\delta f)=[D_M,f]=\pink{-i}\sum_\mu \gamma^\mu\partial_\mu f$,
where $\gamma^\mu$ runs over the $\gamma$-matrices
of the spin bundle. If we consider
$\omega=g(\delta f)- (\delta f)g$, then
$\pi(\omega)=\pink{-i}\sum_\mu (g\partial_\mu f  - \partial_\mu 
fg)\gamma^\mu=0$ because the functions
$g$ and $\partial_\mu f$ commute. However,
$\delta\omega=(\delta g)(\delta f)+(\delta f)(\delta g)$ 
by the graded Leibniz rule and
$\pi(\delta\omega)$ is generally not zero because
\begin{eqnarray*}
\pi(\delta\omega) &=& 
\pink{-}\sum_{\mu\nu} \partial_\mu f \partial_\nu g
 (\gamma^\nu\gamma^\mu+ \gamma^\mu\gamma^\nu)
\\&=& \pink{-}2 \sum_{\mu\nu}  g^{\mu\nu} \partial_\mu f \partial_\nu g
\,  \un=
\pink{-}2 (\partial f)\cdot (\partial g)\,\un,
\end{eqnarray*}
where $\un$ is the unit matrix in the spinor \pink{fiber}.

For a general spectral triple, Connes and 
Lott~\cite{Connes-92} remove all the badly-behaving forms
by defining the \emph{junk} $J=J_0+\delta J_0$, where
$J_0=\bigoplus_{n\ge0} J_0^n$ and $J_0^n=\{\omega\in\Omega^n\telque
\pi(\omega)=0\}$. The ideal $J_0$ is the
kernel of $\tilde\pi$ but not the kernel of $\pi$.
The term $\delta J_0$ is needed because $J_0$
is a graded ideal of $\Omega$ but 
not a differential ideal (i.e. $\delta J_0$
is generally not a subset of $J_0$). 
But $J$ is a graded
differential ideal of $\Omega$
because $\delta^2=0$ implies $\delta J=\delta J_0\subset J$
and $\Omega_D=\Omega/J$ is now a
well-defined DGA called the space of
\emph{noncommutative differential forms} of the spectral triple. 
Moreover,\cite{Landi-NCG}
\begin{eqnarray}
\Omega_D &=& \Omega/J \cong 
\bigoplus_{n\ge0}\pi(\Omega^n)/\pi(\delta J_0^{n-1}).
\label{defOmegaD}
\end{eqnarray}
For a spin manifold, $\Omega_D$ is then isomorphic to the
usual space $\Gamma(M,\Lambda T^*M)$ of differential forms on $M$.

Why don't we represent $\Omega$ over $\calH$?  Indeed, 
since $\calA$ is represented over $\calH$, it would be tempting
to represent $\Omega$ over a graded version of $\calH$
(i.e.  a graded vector space $V$ where every $V^n=\calH$)
and to represent $\Omega_D$ as some quotient.
However, in such a picture we would have to
represent $\Omega_D$ over the graded vector space $W$ where
$W^n=\pi(\Omega^n)\calH/\pi(\delta J_0^{n-1})\calH$
and this quotient is
often trivial. For the example of the spin manifold,
we saw that $ (\partial f\cdot\partial g)\,\un$ belongs to
$\pi(\delta J_0^1)$. As a consequence,
$\pi(\delta J_0^1)\calH=\calH$ and $\calM^2=\{0\}$.

Our purpose is now to extend $\Omega_D$ by
$\calM_D$ in the sense of Eilenberg, where $\calM_D$ is a differential
graded bimodule over $\Omega_D$ naturally defined out of 
the spectral triple data.
This will be done in two steps. We first explain how a 
left-right graded representation of  $\Omega$ can be viewed as 
a left graded representation of a certain algebra $B$. Then we 
take the junk into account, what leads us to quotient $B$ 
by an ideal $K$. We then obtain a graded algebra $B/K$
 which, under some condition,  has the ability to produce 
differential graded $\Omega_D$-bimodules out 
of graded $B$-bimodules.

\subsection{Left and right representations of
$\Omega$ as left representations of $B$}

The only mean at our disposal to produce a right action of $\Omega$ 
is by extending the definition of the map $x\mapsto x^0$ from 
$\pi(A)$ to $\pi(\Omega)$. 
\pink{This extension, uniquely determined by
$\pi(\delta a)^\circ = [D,\pi(a)^\circ]=\delta \pi(a)^\circ$ and
condition~(\ref{aoactbo}), is defined by:}
\begin{eqnarray}
\pi(\omega)^\circ &=& (-1)^{|\omega|(|\omega|+1)/2}
  (\epsilon')^{|\omega|} J \pi(\omega)^\dagger J^{-1},
\label{piomegao}
\end{eqnarray}
where $\epsilon'$ is such that
$JD=\epsilon'DJ$. 
This definition is also compatible with 
\pink{the involution}
$(\delta a)^*=-\delta(a^*)$.~\cite{Connes94}

The receptacle for the objects we need to 
manipulate is the graded
$*$-algebra generated by all the elements 
of the form $\pi(\omega)$ or $\pi(\omega)^\circ$ for 
$\omega\in \Omega$. We call $B=\bigoplus_{n\ge 0} B^n$ 
this algebra, where each $B^n\subset B(\calH)$
and we observe that the grading of $B$ follows from the
grading of $\Omega$:
$|\pi(a)|=|\pi(a)^\circ|=0$, 
$|\pi(\delta a)|=|\pi(\delta a)^\circ|=1$.

Consider now a graded left-representation $\calM$ of $B$.
Then $\calM$ is automatically a graded left and right 
representation of
$\Omega$ with the following actions for homogeneous 
elements $\omega\in \Omega$ and $m\in\calM$:

\begin{eqnarray*}
\omega\triangleright m &=& \pi(\omega)m\cr
m\triangleleft \omega &=& (-1)^{|\omega||m|}\pi(\omega)^\circ m.
\end{eqnarray*}

Let us check that $\triangleleft$ indeed defines a right action:

\begin{eqnarray*}
(m\triangleleft \omega)\triangleleft \omega' &=& (-1)^{|\omega'|(|m|+|\omega|)+|\omega||m|}\pi(\omega')^\circ \pi(\omega)^\circ m\cr
&=&(-1)^{(|\omega'|+|\omega|)|m|}\pi(\omega\omega')^\circ m\cr
&=&m\triangleleft (\omega\omega').
\end{eqnarray*}

\subsection{Bimodule over $\Omega_D$}

Thus $\calM$ is a graded left and right representation $\Omega$ but 
it is not a bimodule: the left and right actions are not compatible 
in general. Moreover, we saw that the elements of $\pi(\Omega)$ 
cannot be properly identified with differential forms 
which are given by the quotient of Eq.~(\ref{defOmegaD}).
Because of the isomorphism described by Eq.~(\ref{defOmegaD}),
we can consider an element 
of $\Omega_D$ from two equivalent points of views:
either as a class $[\omega]$ of 
universal differential forms $\omega$, 
such that $[\omega]=[\omega']$ iff there is \pink{an}
$\eta$ and a $\rho$ in $J_0$ such that
$\omega'=\omega + \rho + \delta \eta$,
or as a class $\langle\alpha\rangle$ of elements of $\pi(\Omega)$ such
that $\langle \alpha\rangle=\langle\alpha'\rangle$ iff there is an element
$\eta$ of $J_0$ such that $\alpha'=\eta+\pi(\delta\eta)$.
Since $J$ is an ideal of $\Omega$ and
$\pi(\delta J_0)$ is an ideal of $\pi(\Omega)$, the product
$[\omega][\omega']=[\omega\omega']$ or
$\langle \alpha\rangle\langle \alpha'\rangle=
\langle\alpha\alpha'\rangle$ are well defined and
$[\omega\omega']=\langle \alpha\alpha'\rangle$ if
$\alpha=\pi(\omega)$ and $\alpha'=\pi(\omega')$.
Moreover, $\delta[\omega]=[\delta\omega]=\langle\pi(\delta\omega)\rangle$ 
is now a well-defined differential on $\Omega_D$.

Here, $\Omega_D$ was built as the quotient of
$\pi(\Omega)$ by the ideal $\pi(\delta J_0)$.
Similarly, we can define $\Omega_D^\circ$ as the
graded quotient of $\pi(\Omega)^\circ$ by 
$\pi(\delta J_0)^\circ$. More precisely,
we define $\Omega_D^\circ$ as the set
of classes $\langle \alpha^\circ\rangle$ where
$\langle\alpha^\circ\rangle =\langle\beta^\circ\rangle$ iff there is an
$\eta\in J_0$ such that $\beta^\circ=\alpha^\circ + (\delta \eta)^\circ$.
This defines a map $\Omega_D\to \Omega_D^\circ$ by
$\langle\alpha\rangle^\circ=\langle\alpha^\circ\rangle$.
Note that the product $\alpha^\circ\beta^\circ$ is
well defined as a product in $B^\infty(\calH)$.
Since $(\delta J_0)^\circ$ is an ideal in
$\pi(\Omega)^\circ$ we can define similarly
\begin{eqnarray*}
\langle\alpha\beta\rangle^\circ &=&
   (-1)^{|\alpha||\beta|} \langle\beta^\circ\alpha^\circ\rangle=
   (-1)^{|\alpha||\beta|} \langle\beta\rangle^\circ\langle\alpha\rangle^\circ,
\end{eqnarray*}
where we used the fact that
$(\alpha\beta)^\circ=(-1)^{|\alpha||\beta|} \beta^\circ\alpha^\circ$
in $B^\infty(\calH)$.
Finally, the differential on $\Omega_D$ is compatible
with ${}^\circ$ in the sense that
$\delta\langle\alpha^\circ\rangle=\delta\langle\alpha\rangle^\circ=\langle\delta\alpha\rangle^\circ$
is well defined. 
Thus, the compatibility equations~(\ref{aoactbo}) and
(\ref{deltaao}) are satisfied.

To complete the conditions on ${}^\circ$ we still have
to satisfy Eq.(\ref{aoactb}). For this, we first
must define the products
$\langle\alpha\rangle^\circ\langle\beta\rangle$ and $\langle\beta\rangle\langle\alpha\rangle^\circ$.
Since $\alpha$ and $\beta^\circ$ are elements of $B^\infty(\calH)$,
the product $\alpha\beta^\circ$ is well defined in
$B^\infty(\calH)$.
Let us consider $\alpha'=\alpha+\pi(\delta\eta)$
and $\beta'=\beta+\pi(\delta\zeta)$. Then
\begin{eqnarray*}
\alpha'(\beta')^\circ &=&
\alpha\beta^\circ + \pi(\delta\eta)\beta^\circ
+ \alpha(\delta\zeta)^\circ + \pi(\delta\eta)(\delta\zeta)^\circ.
\end{eqnarray*}
Since we need $\langle\alpha'(\beta')^\circ\rangle=\langle\alpha\beta^\circ\rangle$
for the product $\langle\alpha\rangle\langle\beta\rangle^\circ$ to be well defined,
all the terms following $\alpha\beta^\circ$ must belong to
an ideal $K$. By multiplying with other elements of
$\pi(\Omega)$ or $\pi(\Omega)^\circ$, we see that
$K$ is the graded ideal generated by $\pi(\delta J_0)+
\pi(\delta J_0)^\circ$ in the graded algebra $B$.
In $B/K$, the products
$\langle\alpha\rangle\langle\beta^\circ\rangle$  and 
$\langle\beta\rangle^\circ \langle\alpha\rangle$
are now well defined. Moreover, and this is an
important check, if $b^\circ\in \calA$
(more precisely, if, for every $b\in \calA$, there
is a $c\in \calA$ such that  $\pi(c)=\pi(b)^\circ$),
then $B=\pi(\Omega)$, $K=J$ and $B/K=\Omega_D$.
Note that this is the case of the
canonical spectral triple of a spin manifold
because $f^\circ=f$.

Since junk forms act on the right as well as on the left, 
$K$ is a graded  ideal of $B$ and
$B/K$ is a graded algebra. 
In the following, we shall use the representation
$\calM_D=B/K$ but more generally,
any left representation $\calM$ of $B$ gives 
rise to a left representation $\calM_D=(B/K)\otimes_B \calM$ of
$B/K$ by extension of scalars, and what is more, $\calM_D$ 
is automatically a left and right representation of $\Omega_D$. 
The left and right actions of $\Omega_D$ on $\calM_D$ 
are explicitly given by:

\begin{eqnarray*}
[\omega]\triangleright m &=& \langle\pi(\omega)\rangle m,\\
m\triangleleft [\omega] &=& (-1)^{|\omega||m|}
   \langle \pi(\omega)^\circ\rangle m.
\end{eqnarray*}

These actions are obviously well-defined since the difference 
between two representatives of $[\omega]\in \Omega_D$ 
belongs to $K$.
The two actions will be compatible, that is 
$([\omega]\triangleright m)\triangleleft [\omega']=
[\omega]\triangleright (m\triangleleft [\omega'])$ if and only if

\begin{eqnarray}
\pi(\omega)\pi(\omega')^\circ -(-1)^{|\omega||\omega'|}
  \pi(\omega')^\circ\pi(\omega)=0\mod K,
\end{eqnarray}
for all homogeneous $\omega,\omega'\in\Omega$. 
Since $\Omega$ is generated as an algebra by elements of 
degree $0$ and $1$, it is equivalent to require that the 
usual order $0$ and order $1$ condition of spectral triples 
hold modulo $K$, which they obviously do, and that moreover:

\begin{eqnarray}
\pi(\delta a)\pi(\delta b)^\circ + 
\pi(\delta b)^\circ \pi(\delta a) &=& 0\mod K.
\label{ordertwo}
\end{eqnarray}

\section{Application to the Standard Model}
\pink{As we} saw, condition~(\ref{aoactb}):
$\langle \alpha\rangle^\circ\langle\beta\rangle = 
(-1)^{|\alpha||\beta|} 
\langle\beta\rangle\langle\alpha\rangle^\circ$ is equivalent to
the four equations
\begin{eqnarray*}
\pi(a) \pi(b)^\circ-\pi(b)^\circ \pi(a) &=& 0,\\
{[}D,\pi(a){]}\pi(b)^\circ-\pi(b)^\circ[D,\pi(a)]&=& 0,\\
{[}D,\pi(a)^\circ{]}\pi(b)-\pi(b)[D,\pi(a)^\circ]&=&0,\\
{[}D,\pi(a){]}[D,\pi(b)^\circ] +
[D,\pi(b)^\circ] [D,\pi(a)]&=&0\mod K.
\end{eqnarray*}
The first equation is satisfied because it is the condition of order zero,
the second equation is the condition of order one,
the third equation is a consequence of the condition of order one,
the fourth equation is called the \emph{condition of order two}. 
It is new and we investigate 
it for the spectral triple 
$(\calA,\calH,D,J,\gamma)$
of the Standard Model, which is
the tensor product of 
$(C^\infty(M),L^2(M,S),D_M,J_M,\gamma^5)$ and
$(\calA_F,\calH_F,D_F,J_F,\gamma_F)$
and where
$D=D_M\otimes \id + \gamma^5\otimes D_F$.
Let us first consider these
four conditions in a tensor product of 
general even spectral triples.

\subsection{Tensor product of even spectral triples}
\label{tensor-sect}
If $(\calA_1,\calH_1,D_1,J_1,\gamma_1)$ 
and
$(\calA_2,\calH_2,D_2,J_2,\gamma_2)$ are even
spectral triples with representation maps
$\pi_1$ and $\pi_2$, then their tensor product is
defined by
$\calA=\calA_1\otimes\calA_2$,
$\calH=\calH_1\otimes\calH_2$,
$D=D_1\otimes \id_2 + \gamma_1\otimes D_2$ and
$\gamma=\gamma_1\otimes
\gamma_2$.\cite{Vanhecke-99,Dabrowski-11,Cacic-13}
If the KO-dimension of the first spectral triple
is 4, then $J=J_1\otimes J_2$.

The conditions of order zero and one hold for this
tensor product~\cite{Dabrowski-11} but we must
investigate the condition of order two.
Let $a=a_1\otimes a_2$ and $b=b_1\otimes b_2$,
by using the fact that $\gamma_1$ is unitary, self-adjoint,
commutes with all elements of $\calA_1$ and anticommutes
with $D_1$ we obtain 
\begin{eqnarray*}
\{[D,a],[D,b^\circ]\}
   &=& \{[D_1,a_1],[D_1,b_1^\circ]\}\otimes a_2  b^\circ_2
\\&&
    +
    a_1 b_1^\circ\otimes \{[D_2,a_2],[D_2,b^\circ_2]\}
\end{eqnarray*}
and the condition of order two means that this
anticommutator
must belong to the junk $K$ of the tensor product.

In general,~\cite{Madore} the universal DGA
$\Omega$ built from $\calA_1\otimes\calA_2$ 
is different from 
the DGA $\Omega_1\otimes\Omega_2$
and the expression of $\Omega_D$ in terms
of $\Omega_{D1}$ and $\Omega_{D2}$ is
rather intricate.\cite{Kalau-95}

However, when $\calA_1=C^\infty(M)$ and
$\calA_2=\calA_F$ the situation is simpler
and it can be shown that~\cite{Martin-98}
\begin{eqnarray*}
\pi_\otimes (\delta J_0^1) &=& 
\pi_M(\delta J_{0M}^1) \otimes \pi(\calA_F) + C^\infty(M)
  \otimes \pi(\delta  J_{0F}^1)
\end{eqnarray*}
where $\pi_\otimes=\pi_M\otimes \pi$.

\pink{Since there is no element of degree zero 
or one in $\pi(J)$}, the space $K^2$ of elements of degree two
in the junk of the tensor product is
\begin{eqnarray*}
K^2 &=& \pi_\otimes(\calA)^\circ \pi_\otimes(\delta J_0^1)+
\pi_\otimes(\calA) \pi_\otimes(\delta J_0^1)^\circ.
\end{eqnarray*}
More precisely,
\begin{eqnarray*}
K^2 &=& \pi_M(\delta J_{0M}^1) \otimes \pi(\calA_F)\pi(\calA_F)^\circ
\\&&
+ C^\infty(M)
  \otimes \pi(\delta  J_{0F}^1) \pi(\calA_F)^\circ
\\&&
+ C^\infty(M)
  \otimes \pi(\delta  J_{0F}^1)^\circ \pi(\calA_F),
\end{eqnarray*}
which must be completed by 
$\pi_M(\delta J_{0M}^1)=C^\infty(M) \un$,
where $\un$ is the identity of the spinor bundle.

To summarize this discussion, the condition of order two
is satisfied for the tensor product $\calA=C^\infty(M)\otimes\calA_F$
if and only if it is satisfied for $C^\infty(M)$ and
the anticommutator
$\{[D_F,a],[D_F,b^\circ]\}$ belongs to $\pi(\calA_F)\pi(\calA_F)^\circ+
\pi(\delta  J_{0F}^1) \pi(\calA_F)^\circ
+ \pi(\calA_F)\pi(\delta  J_{0F}^1)^\circ$.
This is what we are going to check in the next sections.

\subsection{Spin manifold}
For the spectral triple of a spin manifold,
$\pi(f)^\circ=J f^*J^{-1}=fJJ^{-1}=f$.
Thus, $\pi(\calA)^\circ=\pi(\calA)$ and
the right action of $C^\infty(M)$ over 
$C^\infty(M,S)$ is the same as the
left action. As a consequence,
$\calM_{D_M}=\Omega_{D_M}$ is obviously a
differential graded bimodule over itself
and we do not need to check the condition
of order two. Let us do it anyway by calculating
\begin{eqnarray*}
[D_M,f][D_M,g]+[D_M,g][D_M,f] &=& 
\pi(\delta\omega),
\end{eqnarray*}
where $\omega=g(\delta f)- (\delta f)g$ 
was defined in section~\ref{ncdiform}.
Since $\pi(\omega)=0$,  then
$\pi(\delta\omega)\in\pi(\delta J_0)$ and 
$\{[D_M,f],[D_M,g]\}$ indeed belongs to the junk. 

\subsection{The finite spectral triple}
\label{finite-sect}

\pink{Since we consider a single generation,
the 32-dimensional Hilbert space $\calH_F$ 
can be split into four 8-dimensional subspaces
$\calH_F=\calH_R\oplus \calH_L\oplus\calH_{\overline{R}}
\oplus \calH_{\overline{L}}$, where 
$\calH_R$ describes the right-handed particles,
$\calH_L$  the left-handed particles
and $\calH_{\overline{R}}$
 and $\calH_{\overline{L}}$ their antiparticles.
}
An element of the finite algebra $\calA_F$
is parametrized by a complex number $\lambda$,
a quaternion written as a pair of complex 
numbers $(\alpha,\beta)$ and a
3x3 matrix $\mu$. 
Its representation over $\calH_F$ is:
\begin{eqnarray*}
  \pi(a) &=& \left(\begin{array}{cccc}
      \mathsf{A} & 0 &  0 &  0 \\
      0 & \mathsf{B}  & 0 & 0 \\
      0 & 0 & \mathsf{C} & 0 \\
      0 & 0 & 0 & \mathsf{C} 
    \end{array}\right),
\end{eqnarray*}
and
\begin{eqnarray*}
  \pi(a)^\circ &=& \left(\begin{array}{cccc}
      \mathsf{C}^T & 0 &  0 &  0 \\
      0 & \mathsf{C}^T  & 0 & 0 \\
      0 & 0 & \mathsf{A} & 0 \\
      0 & 0 & 0 & \mathsf{B}^T 
    \end{array}\right),
\end{eqnarray*}
with
\begin{eqnarray*}
  \mathsf{A} &=& \left(\begin{array}{cccccccc}
      \lambda & 0 & 0 & 0 & 0 & 0 & 0 & 0\\
      0 & \bar{\lambda} & 0 & 0 & 0 & 0 & 0 & 0\\
      0 & 0 & \lambda  & 0 & 0 & 0 & 0 & 0\\
      0 & 0 & 0 & \lambda & 0 & 0 & 0 & 0 \\
      0 & 0 & 0 & 0 & \lambda  & 0 & 0 & 0 \\
      0 & 0 & 0 & 0 & 0 & \bar{\lambda} & 0 & 0 \\
      0 & 0 & 0 & 0 & 0 & 0 & \bar{\lambda}  & 0 \\
      0 & 0 & 0 & 0 & 0 & 0 & 0 & \bar{\lambda}
    \end{array}\right),
\end{eqnarray*}
\begin{eqnarray*}
  \mathsf{B} &=& \left(\begin{array}{cccccccc}
      \alpha & \beta & 0 & 0 & 0 & 0 & 0 & 0\\
     -\bar{\beta} & \bar{\alpha} & 0 & 0 & 0 & 0 & 0 & 0\\
      0 & 0 & \alpha &  0 & 0 & \beta & 0 & 0\\
      0 & 0 & 0 & \alpha &  0 & 0 & \beta & 0\\
      0 & 0 & 0 & 0 & \alpha &  0 & 0 & \beta\\
      0 & 0 & -\bar{\beta} &  0 & 0 & \bar{\alpha} & 0 & 0\\
      0 & 0 & 0 & -\bar{\beta} &  0 & 0 & \bar{\alpha} & 0\\
      0 & 0 & 0 & 0 & -\bar{\beta} &  0 & 0 & \bar{\alpha}
    \end{array}\right),
\end{eqnarray*}
\begin{eqnarray*}
  \mathsf{C} &=& \left(\begin{array}{cccccccc}
      \lambda & 0 & 0 & 0 & 0 & 0 & 0 & 0\\
      0  & \lambda & 0 & 0 & 0 & 0 & 0 & 0\\
      0 & 0 & \mu_{11} &  \mu_{12} & \mu_{13} & 0 & 0 & 0\\
      0 & 0 & \mu_{21} &  \mu_{22} & \mu_{23} & 0 & 0 & 0\\
      0 & 0 & \mu_{31} &  \mu_{32} & \mu_{33} & 0 & 0 & 0\\
      0 & 0 & 0 & 0 & 0 & \mu_{11} &  \mu_{12} & \mu_{13}\\
      0 & 0 & 0 & 0 & 0 & \mu_{21} &  \mu_{22} & \mu_{23}\\
      0 & 0 & 0 & 0 & 0 & \mu_{31} &  \mu_{32} & \mu_{33}
    \end{array}\right).
\end{eqnarray*}
The antilinear real structure $J$ acts by
\begin{eqnarray*}
J\Big(\sum_{i=1}^{32} v_i e_i\Big) &=&
\sum_{i=1}^{16} \overline{v_{i+16}} e_i
+\sum_{i=17}^{32} \overline{v_{i-16}} e_i.
\end{eqnarray*}
where $(e_1,\dots,e_{32})$ is a basis of $\calH_F$.
\pink{The chirality operator is}
\begin{eqnarray*}
  \gamma &=& \left(\begin{array}{cccc}
      -\un_8 & 0 &  0 &  0 \\
      0 & \un_8  & 0 & 0 \\
      0 & 0 & \un_8 & 0 \\
      0 & 0 & 0 &  -\un_8
    \end{array}\right).
\end{eqnarray*}
The relations $D_F^\dagger=D_F$, 
$D_F J= J D_F$, $D_F\gamma=-\gamma D_F$ and 
the condition of order one imply the following
form for the Dirac operator:
\begin{eqnarray*}
  D_F &=& \left(\begin{array}{cccc}
      0 & \mathsf{Y}^\dagger &  \mathsf{M}^\dagger &  0 \\
      \mathsf{Y} & 0 & 0 & 0 \\
      \mathsf{M} &  0 & 0 &  \mathsf{Y}^T \\
      0 & 0 & \overline{\mathsf{Y}} &  0
    \end{array}\right),
\end{eqnarray*}
with the Yukawa matrix
\begin{eqnarray*}
  \mathsf{Y} &=& \left(\begin{array}{cccccccc}
      l_{11} & l_{12} & 0 & 0 & 0 & 0 & 0 & 0\\
      l_{21} & l_{22} & 0 & 0 & 0 & 0 & 0 & 0\\
      0 & 0 & q_{11} &  0 & 0 & q_{12} & 0 & 0\\
      0 & 0 & 0 & q_{11} &  0 & 0 & q_{12} & 0\\
      0 & 0 & 0 & 0 & q_{11} &  0 & 0 & q_{12}\\
      0 & 0 & q_{21} &  0 & 0 & q_{22} & 0 & 0\\
      0 & 0 & 0 & q_{21} &  0 & 0 & q_{22} & 0\\
      0 & 0 & 0 & 0 & q_{21} &  0 & 0 & q_{22}\\
    \end{array}\right),
\end{eqnarray*}
where $l_{ij}$ stands for $y_{l,ij}$ and
$q_{ij}$ for $y_{q,ij}$ in the notation
used by Boyle and Farnsworth and the mass matrix
\begin{eqnarray*}
  \mathsf{M} &=& \left(\begin{array}{cccccccc}
      a & b & c_1 & c_2 & c_3 & 0 & 0 & 0\\
      b & 0 & d_1 & d_2 & d_3 & 0 & 0 & 0\\
      c_1 & d_1 & 0 & 0 & 0 & 0 & 0 & 0\\
      c_2 & d_2 & 0 & 0 & 0 & 0 & 0 & 0\\
      c_3 & d_3 & 0 & 0 & 0 & 0 & 0 & 0\\
       0 & 0 & 0 & 0 & 0 & 0 & 0 & 0 \\
       0 & 0 & 0 & 0 & 0 & 0 & 0 & 0 \\
       0 & 0 & 0 & 0 & 0 & 0 & 0 & 0
    \end{array}\right),
\end{eqnarray*}
where $a$ corresponds to the mass of the right-handed
neutrino and the other parameters are eliminated
by the condition of massless photon.
For notational convenience we write
$\bfc=(c_1,c_2,c_3)$ and $\bfd=(d_1,d_2,d_3)$.

\begin{figure}
\begin{center}
\includegraphics[width=7.0cm]{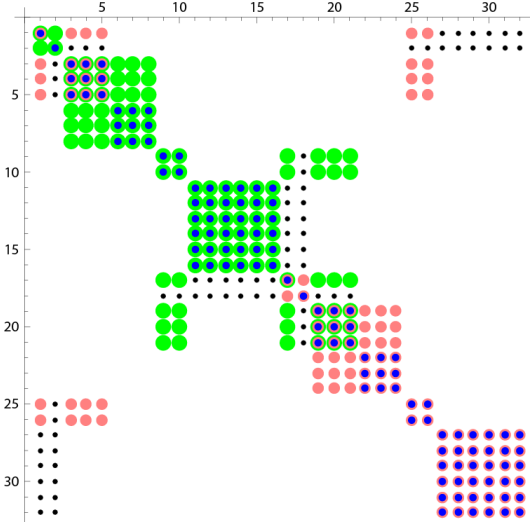}
\caption{Each dot at position $(i,j)$ corresponds to
a generally non-zero element at line $i$ and column $j$.
The elements of the condition of order two 
$\{[D,a],[D,b^\circ]\}$ are black dots.
The other dots describe the junk.
The green dots correspond to $[D,a][D,b]c^\circ$,
the pink dots to $[D,a^\circ][D,b^\circ]c$ and
the blue dots to $ab^\circ$.
Note that the \pink{non-zero} matrix elements of the junk and of
the second-order condition do not \pink{overlap}.
  \label{figjunk}}
\end{center}
\end{figure}

We must check that $\{[D_F,a],[D_F,b^\circ]\}=0$
up to the junk. Since the condition of order two is
of degree two, we need the junk of degree two, which 
was determined in section~\ref{tensor-sect}.
We do not try to determine $K^{2}$ more explicitly.
We calculate the $32\times 32$ matrices
$m=\pi(\delta a') \pi(\delta b')\pi(c)^{\pink{o}}$,
$n=\pi(\delta a)^\circ\pi(\delta b)^\circ \pi(c')$
and 
$p=\pink{\pi(e')\pi(e)^\circ}$
for generic $a$, $b$, $c$, $a'$, $b'$, $c'$,
$e$ and $e'$ in $\calA_F$ and we notice that there are 820
pairs of indices $(k,l)$ such that
$m_{kl}=n_{kl}=p_{kl}=0$ (see Fig.~\ref{figjunk}).

Among the pairs of indices where the junk is zero, 
68 of them correspond to matrix elements of 
$\{[D_F,a],[D_F,b^\circ]\}$ that are not generically zero
 (see Fig.~\ref{figjunk}).
Since they cannot be compensated by the junk, 
the condition of order two implies that these 68 matrix
elements must be equal to zero. 
Because of the symmetry generated by the adjoint ${}^\dagger$
and the ${}^\circ$ operations, this
gives 17 different equations
that can be grouped into three systems.
Let the elements $a$ and $b$ of the algebra be parametrized
by $\lambda,\alpha,\beta,\mu$ and
$\lambda',\alpha',\beta',\mu'$, respectively.
The first system (of two equations):
\begin{eqnarray*}
\big((\alpha - \lambda)y_{l,11} + \beta y_{l,21}\big)
  \bar{b} (\lambda'-\bar{\lambda'}) &=& 0,\\
\big((\bar{\alpha} - \lambda)y_{l,21} - \bar{\beta} y_{l,11}\big)
  \bar{b} (\lambda'-\bar{\lambda'}) &=& 0.
\end{eqnarray*}
is solved by
either $b=0$ or $y_{l,11}=y_{l,21}=0$ in the lepton Yukawa matrix.
The second system (of 12 equations):
\begin{eqnarray*}
\big((\alpha - \lambda)y_{l,11} + \beta y_{l,21}\big)
  (\sum_{i=1}^3 \bar{d}_i \mu'_{ij} - \bar{\lambda}' \bar{d}_j) &=& 0,\\
\big((\bar{\alpha} - \lambda)y_{l,21} - \bar{\beta} y_{l,11}\big)
  (\sum_{i=1}^3 \bar{d}_i \mu'_{ij} - \bar{\lambda}' \bar{d}_j) &=& 0,\\
\big((\alpha - \lambda)y_{l,11} + \beta y_{l,21}\big)
  (\sum_{i=1}^3 \bar{c}_i \mu'_{ij} - \lambda' \bar{c}_j) &=& 0,\\
\big((\bar{\alpha} - \lambda)y_{l,21} - \bar{\beta} y_{l,11}\big)
  (\sum_{i=1}^3 \bar{c}_i \mu'_{ij} - \lambda' \bar{c}_j) &=& 0,
\end{eqnarray*}
where $j=1,2,3$,
is solved by $\bfc=\bfd=0$ or 
$y_{q,11}=y_{q,21}=0$ in the quark Yukawa matrix.
The third system (of three equations):
\begin{eqnarray*}
  \bar{b} (\lambda'-\bar{\lambda'}) 
  (\sum_{i=1}^3 c_i \mu_{ij} - \lambda c_j) &=& 0,
\end{eqnarray*}
where $j=1,2,3$,  is solved by either $b=0$ or $\bfc=0$.
By putting these solutions together we recover exactly the
four solutions found by Boyle and Farnsworth:
(i) $b=\bfc=\bfd=0$;
(ii) $b=y_{q,11}=y_{q,21}=0$;
(iii) $y_{l,11}=y_{l,21}=\bfc=\bfd=0$;
(iv) $y_{l,11}=y_{l,21}=y_{q,11}=y_{q,21}=\bfc=0$.
Three of these four solutions are not physically acceptable
because they \pink{correspond to Yukawa matrices having
a zero column}.
The remaining solution (i) is precisely the result of the
condition of zero photon mass.

Note that we solved the anticommutator equation
$\{[D_F,a],[D_F,b^\circ]\}=0$ up to the junk
while Boyle and Farnsworth solve the commutator equation
$\big[[D_F,a],[D_F,b^\circ]\big]=0$ 
without junk condition and it may seem surprising that
we find the same solutions.
This is explained by the fact that
\pink{the non-zero elements of 
$[D_F,a][D_F,b^\circ]$ and $[D_F,b^\circ][D_F,a]$
do not overlap. Thus, to cancel the commutator
or the anticommutator we need
$[D_F,a][D_F,b^\circ]=[D_F,b^\circ][D_F,a]=0$.}
Moreover, we do not need to determine the
junk more precisely since we already
have $\{[D_F,a],[D_F,b^\circ]\}=0$ without
junk condition.

\section{Lorentzian Standard Model}
We show now that the second-order condition is also
compatible with a \pink{Lorentzian} spectral triple of the 
Standard Model \pink{on a Lorentzian spin manifold}.

\subsection{Lorentzian spectral triples} 
We describe the main aspects of a Lorentzian spectral
triple, slightly changing the notation to make it
more compatible with the physics literature.
A \emph{Krein space} is a Hilbert space $\calH$ equipped 
with a self-adjoint unitary operator $\calJ=\calJ^\dagger=\calJ^{-1}$
called a \emph{fundamental symmetry}.
The \emph{Krein-adjoint} of an operator 
$T$ on $\calH$ is $T^\times = \calJ T^\dagger \calJ$.
An operator $T$ is \emph{Krein-anti-self-adjoint} if
$T^\times=-T$.

\pink{
By putting together the works of Strohmaier~\cite{Strohmaier-06}
and Paschke and Sitarz~\cite{Paschke-06} we can propose
the following definition.
}
A \emph{real even Lorentzian spectral triple} consists of
(i) a $*$-algebra $\calA$; (ii) a Krein space 
$(\calH,\calJ)$ where every $a\in\calA$ is represented by a bounded
operator $\pi(a)$ such that $\pi(a^*)=\pi(a)^\times$
and $[\calJ,\pi(a)]=0$;
(iii) a Krein-anti-self-adjoint operator $D$ such that
$[D,\pi(a)]$ is bounded for all $a\in \calA$;
(iv) a self-adjoint unitary operator \pink{$\gamma$} that commutes
with $\calA$ and anticommutes with $D$ and $\calJ$;
(v) an antilinear unitary operator $J$ such that
$J^2=\epsilon \un$, $JD=\epsilon'DJ$, $J\gamma=\epsilon''\gamma J$
and $J\calJ=\pm\calJ J$, where $(\epsilon,\epsilon',\epsilon'')$
depends on the $KO$-dimension as for the usual
spectral triples.~\cite{Connes-Marcolli}
\pink{Note that the interplay between fundamental symmetry,
real structure and chirality was also discussed
in studies of topological insulators.\cite{Schulz-13}}

As compared to other definitions of a Lorentzian spectral
triple,\cite{Strohmaier-06,Franco-13,Dungen-15}
we choose $D$ to be Krein-anti-self-adjoint because
the standard Dirac operator is so and we do not need to modify
the $\epsilon$-table giving $(\epsilon,\epsilon',\epsilon'')$
as a function of the $KO$-dimension. 

\pink{
Since $\calJ$ commutes with $\pi(\calA)$ we have
$\pi(a^*)=\pi(a)^\dagger=\pi(a)^\times$.
The space $\Omega_D$
being a $*$-algebra, we need a representation of 
$(\delta a)^*$ compatible 
$(\delta a)^* = - \delta (a^*)$.\cite{Connes94}
The Krein-anti-self-adjointness of the physical
Dirac operator $D$ implies that
$[D,\pi(a)]^\times=[D,\pi(a^*)]$. Therefore,
if we represent $\delta a$ by
$\pi(\delta a)=[iD,\pi(a)]$
we have $\pi(\delta a)^\times =-\pi(\delta a^*)$.
As a consequence, the representation of $(\delta a)^*$
is $\pi(\delta a)^\times$ and the representation
of $a_0 (\delta a_1)\dots(\delta a_n)$ is 
$\pi(a_0)[iD,\pi(a_1)]\dots [iD,\pi(a_n)]$.
The junk is built as in the standard 
case.~\cite{Strohmaier-06}
Moreover, to ensure the validity of Eq.~(\ref{deltaao}),
$\pi(\omega)^\circ$ is given by
formula~(\ref{piomegao}) where 
$\pi(\omega)^\dagger$ is replaced by $\pi(\omega)^\times$.
}
\pink{The} functional properties required for the Dirac
operator are discussed in the
literature.\cite{Strohmaier-06,Dungen-13,Dungen-15}

\pink{By following the same reasoning as for}
a standard spectral triple, 
a real Lorentzian spectral triple should satisfy the 
order-zero condition $[\pi(a),\pi(b)^\circ]=0$,
the order-one conditions
$[[D,\pi(a)],\pi(b)^\circ]=0$,
$[\pi(a),[D,\pi(b)^\circ]]=0$
and the order-two condition
$\{[D,\pi(a)],[D,\pi(b)^\circ]\}=0$.

\subsection{Spectral triple of a Lorentzian spin manifold}
We consider a four-dimensional smooth Lorentzian 
spin manifold $M$ and we
choose the metric signature $(-,+,+,+)$ with $p=3$
positive signs and $q=1$ negative sign  because:
(i) it corresponds to the spectral triple of a spin manifold
with $KO$-dimension $p-q=2$, as advocated by Barrett;\cite{Barrett-07}
(ii) it was argued that this is the only signature
where a neutrinoless double
beta decay can be correctly described.\cite{Berg-01}
Since the following discussion will be local, we can choose
the $\gamma$-matrices to satisfy 
$\{\gamma^\mu,\gamma^\nu\}=g^{\mu\nu}$, where 
$g^{\mu\nu}$ is diagonal with diagonal elements 
$(-1,+1,+1,+1)$.\cite{WeinbergQFT}
We define the helicity operator
$\gamma_M=\gamma^5=-i\gamma^0\gamma^1\gamma^2\gamma^3$,
the Dirac operator
$D_M=-i\gamma^\mu \nabla_\mu$ and the fundamental
symmetry $\calJ_M=\beta=i\gamma^0$, which is used in the 
calculation of expectation
values: $\langle \overline{\psi}|T|\psi\rangle$ with
$|\overline{\psi}\rangle=\beta|\psi\rangle$.
The product $\langle\overline{\psi}|\psi'\rangle$
is called a \emph{Krein product}.
The antilinear map is $J_M=\zeta \calK$, where
$\zeta=\gamma^5 \gamma^2$ and
$\calK$ is the complex conjugate operator.
The operator $J_M$ is
different from the physical charge conjugation
operator $\gamma^2\calK=-\beta\gamma^2\beta\calK$,~\cite{WeinbergQFT}.
In fact there are two 
possible charge conjugation operators
corresponding to two different $\epsilon$-tables,\cite{Dabrowski-11}
$J_M$ is the first one and the physical charge conjugation is
the second one.
It can be checked that all the axioms of a real even
Lorentzian spectral triple hold with these definitions
and that $f^\circ=f$ and $[D_M,f]^\circ=-[D_M,f]$. 
Thus, the order-zero, -one and -two conditions 
are satisfied.

\subsection{The \bleu{Lorentzian} Standard Model}
To describe the Standard Model with Lorentzian
metric, we make the tensor product of the
Lorentzian spectral triple of $M$ 
and the finite spectral triple $\calA_F$
of section~\ref{finite-sect}.
\bleu{The tensor product of pseudo-Riemannian spectral
triples was investigated by van den Dungen.\cite{Dungen-15}}

The grading of the tensor product is
$\gamma=\gamma^5\otimes \gamma_F$,
its Dirac operator
$D=D_M\otimes \id_2 + \gamma^5\otimes D_M$,
its charge conjugation is
$J=J_M\otimes\gamma_F J_F$ because the $KO$-dimensions
 of the first and second spectral triples
are 2 and 6,\cite{Vanhecke-99,Dabrowski-11}
and its fundamental symmetry is
$\calJ=\calJ_M\otimes \un$.
The finite spectral triple cannot be Lorentzian
because this would not be compatible with
the anticommutation of $\gamma$ and $\calJ$.
Moreover, using the finite spectral triple
of section~\ref{finite-sect}
provides the correct fermionic 
Lagrangian in the Lorentzian metric.\cite{Dungen-15}

It can be checked that, with this definition,
the tensor product of spectral triples is indeed
a real even Lorentzian spectral triple of 
$KO$-dimension zero and this solves the 
fermion multiplicity problem.\cite{Barrett-07}
The order-zero and -one conditions hold
by construction.
Moreover, the calculation of section~\ref{tensor-sect}
can be repeated to show that the order-two
condition holds iff 
$\{[D_F,a],[D_F,b^\circ]\}=0$ up to the junk.
Since this was already proved, 
the Lorentzian spectral triple of the Standard Model
satisfies the order-two condition.

\section{Conclusion}
Chamseddine and Connes based \bleu{their} derivation of the Standard
Model on a bimodule over an algebra $\calA$.
Boyle and Farnsworth proposed to use a bimodule
over the universal differential algebra $\Omega$ 
which is physically more satisfactory
because it contains (up to the junk) the gauge fields,
the field intensities, the curvature and the Lagrangian densities.
But their approach was not compatible with the manifold part
of the Standard Model. 

To take into account the differential graded structure
of $\Omega_D$, we built a differential
graded bimodule that takes the junk into account.
The grading transforms the Boyle and Farnsworth  condition 
on the commutator $[\pi(\delta a),\pi(\delta b)^\circ]=0$ into 
a condition on the anticommutator
$\{\pi(\delta a),\pi(\delta b)^\circ\}\in K$,
which is now satisfied for the full 
\pink{Lorentzian} Standard Model
and not only for its finite part.

This indicates that, in a reinterpretation of the
noncommutative geometric approach to field theory,
the differential graded structure of the boson
fields must be accounted for.
This is good news for any future quantization
and renormalization of NCG because the
differential graded structure is also an
essential ingredient of the Becchi-Rouet-Stora-Tyutin
and Batalin-Vilkovisky approaches.

Our differential graded bimodule retains some of the advantages
of the Boyle and Farnsworth approach: (i) it unifies
the conditions of order zero and one and the condition
of massless photon into a single bimodule condition; 
(ii) it can be adapted to non-associative or Lie algebras.

Now we intend to investigate the symmetries of this
approach by using the morphisms defined by 
Eilenberg.~\cite{Eilenberg-48} It will be interesting
to compare these symmetries with the ones found by
Farnsworth and Boyle.~\cite{Farnsworth-15}

We also hope to use our construction for the quantization
of a noncommutative geometric description of the Standard Model
coupled with gravity.

\begin{acknowledgements}
We thank Georges Skandalis and Franciscus Jozef Vanhecke
for very helpful comments.
\end{acknowledgements}



\end{document}